\documentclass[letterpaper, 10 pt, conference]{ieeeconf}

\IEEEoverridecommandlockouts   
\usepackage{graphics} 
\usepackage{epsfig} 
\usepackage{amssymb}
\usepackage{xcolor}

\newcommand{\keypoint}[1]{\textit{#1}\quad}

\usepackage{booktabs}
\usepackage[percent]{overpic}
\newcommand{\ra}[1]{\renewcommand{\arraystretch}{#1}}
\newcommand*\circled[1]{\tikz[baseline=(char.base)]{
            \node[shape=circle,draw,inner sep=1pt] (char) {#1};}}

\usepackage{caption, copyrightbox}

\title{\LARGE \bf
Unlimited Resolution Image Generation with R2D2-GANs*
}

\author{Marija Jegorova$^{1}$, Antti Ilari Karjalainen$^{2}$, Jose Vazquez$^{2}$, Timothy M. Hospedales$^{1}$
\thanks{* This work was supported by SeeByte Ltd}
\thanks{$^{1}$ University of Edinburgh, UK
        {\tt\small m.jegorova@ed.ac.uk, t.hospedales@ed.ac.uk}}%
\thanks{$^{2}$ Seebyte, UK
        {\tt\small antti.karjalainen@seebyte.com, jose.vazquez@seebyte.com}}%
}

\usepackage[belowskip=-6pt,aboveskip=0pt]{caption}
 
\begin{document}

\maketitle{}
\thispagestyle{empty}
\pagestyle{empty}

\begin{abstract}

In this paper we present a novel simulation technique for generating high quality images of any predefined resolution. This method can be used to synthesize sonar scans of size equivalent to those collected during a full-length mission, with across track resolutions of any chosen magnitude. In essence, our model extends Generative Adversarial Networks (GANs) based architecture into a conditional recursive setting, that facilitates the continuity of the generated images. The data produced is continuous, realistically-looking, and can also be generated at least two times faster than the real speed of acquisition for the sonars with higher resolutions, such as EdgeTech. The seabed topography can be fully controlled by the user. The visual assessment tests demonstrate that humans cannot distinguish the simulated images from real. Moreover, experimental results suggest that in the absence of real data the autonomous recognition systems can benefit greatly from training with the synthetic data, produced by the R2D2-GANs. 
\end{abstract}

\section{INTRODUCTION}
\begin{figure}
    \centering
    \vspace{5px}
    \includegraphics[scale = 0.1575]{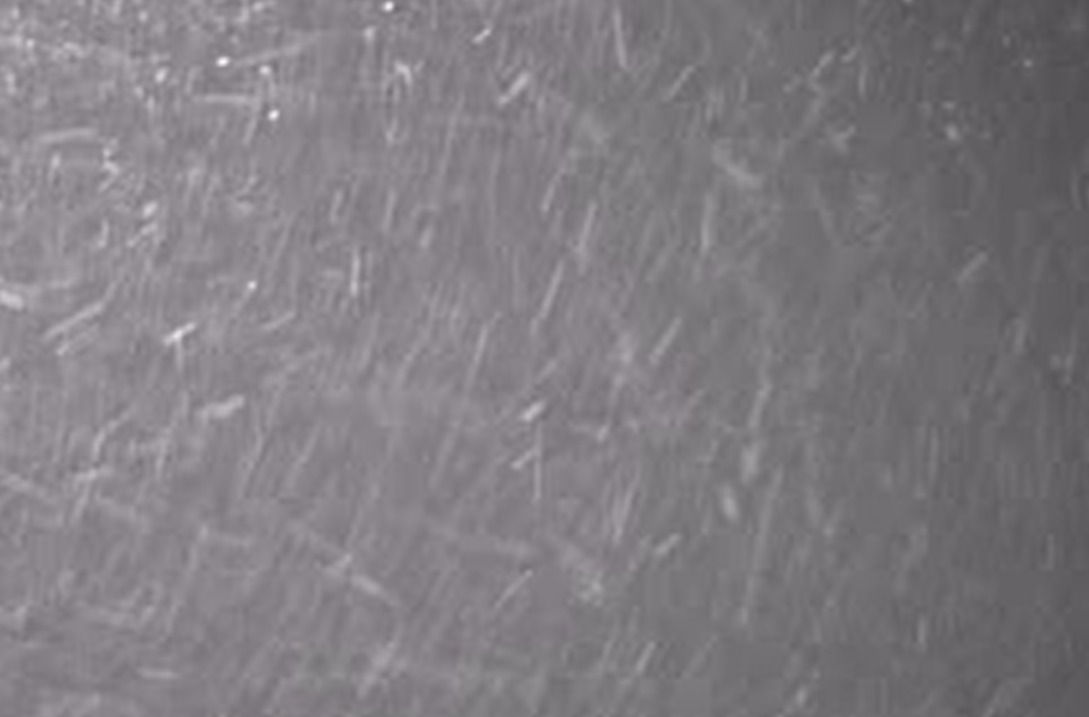}\includegraphics[scale = 0.138]{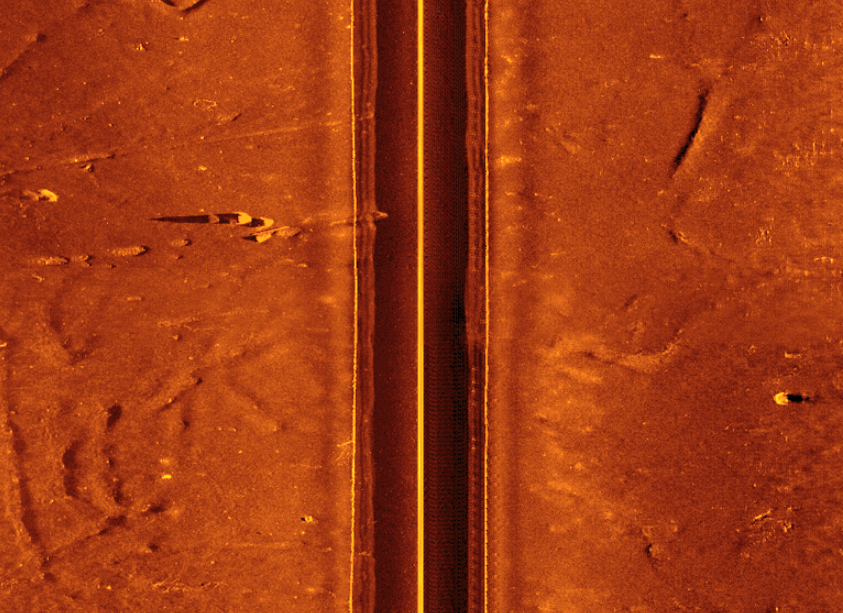} \vspace{0.5px}
    \caption{\textbf{Examples of underwater sensors. Left:} optical camera - an example of visibility in northern seas, this effect is called \textit{marine snow}, it is caused by high density of fine floats. In cases of poor visibility, sonars are often preferred over the cameras as more perceptually robust. \textbf{Right:} sonar side-scan. Port (left) is real, starboard (right) is synthesized.}
    \vspace{-2px}
    \label{visibility}
\end{figure}

Underwater optical visibility is often impaired due to the effect called marine snow, shown in Figure~\ref{visibility} (left), especially in cold seas. Because of that sonars are the primary source of the sensory information for autonomous vehicles operating underwater. The real-life underwater data collection is expensive, time-consuming, and impossible to carry out in certain locations.

However, the vast amounts of data are necessary for automating a number of the data-intensive applications, such as training of autonomous target recognition systems (ATR), as well as training human operators. The data shortage can be addressed by using the limited available data to train a high-quality simulator, capable of producing realistically looking synthetic imagery.

In this paper we propose a technique for synthesis of full-mission-long high-resolution seabed sonar scans, building upon the previously presented method, MC-pix2pix~\cite{MC-pix2pix}, suitable for lower resolution sonars. We suggest to extend the principle of MC-pix2pix further, enabling the resulting generative model not only to preserve all the strengths of its predecessor, but also to generate the data of any chosen high resolution, in principle - any desired resolution. 

The speed of the data generation depends on the hardware, sonar range, and on the required resolution. For instance, for Marine Sonic sonars (512 pixels across track $\times$2 channels) the generation rate is almost 20 times faster than the rate of the real data acquisition. For higher resolution sonars, like EdgeTech (approximately 4620 pixels across track $\times$2 channels), the rate is at least twice faster than the rate of the real data acquisition. These estimates have been acquired with GTX 1080 Ti graphics card (12Gb RAM). 

We call our method double-recursive double-discriminator Generative Adversarial Networks (R2D2-GANs, or ``R2D2" for the sake of conciseness). To our knowledge, this is the first technique capable of adversarial generation of continuous and realistically-looking sonar side-scans of any requested size or resolution.

Potential applications of R2D2 can go far beyond the sonar imagery, as it can produce any type of large resolution imagery, provided a sufficient amount and quality of the initial training examples.

\begin{figure*}
        
        
        \centering
        \begin{overpic}[scale=0.277]{{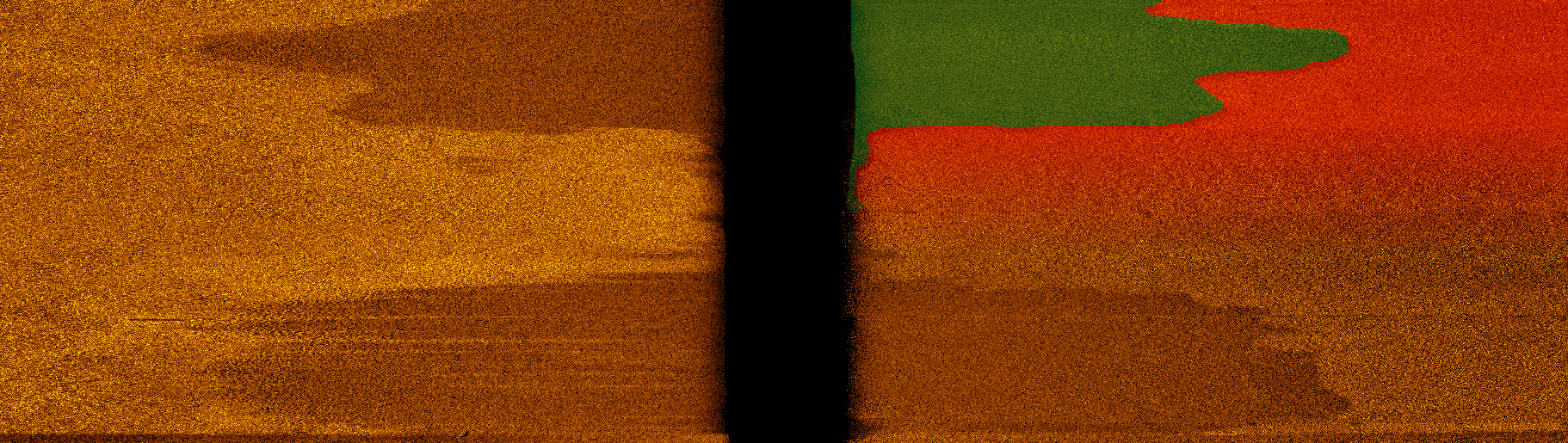}}\put(0, 19.25){\includegraphics[scale=0.075]{MS_square.png}}\end{overpic}\vspace{0px}
        
        
        
        \vspace{0.15cm}
        \begin{overpic}[scale=0.2705]{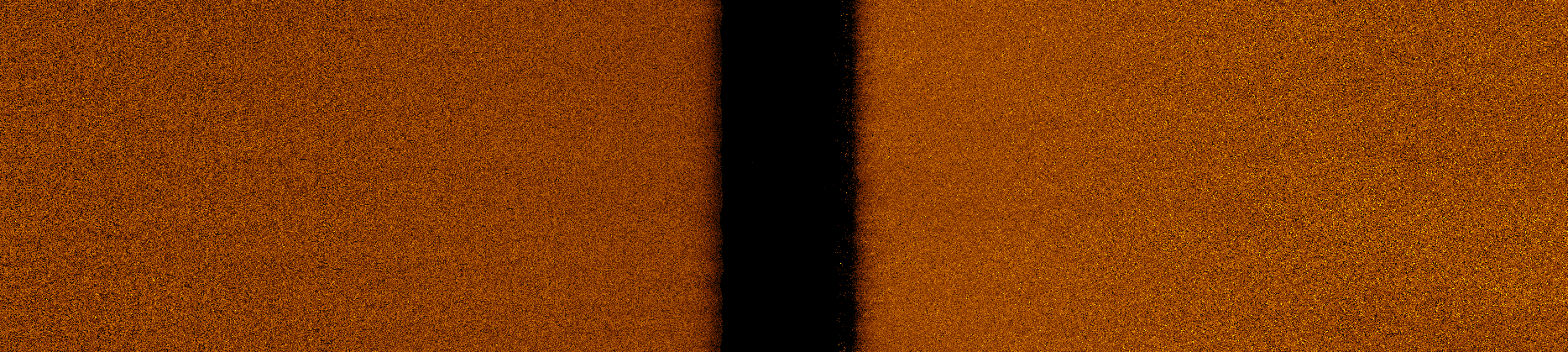}\put(2, 2){\textbf{\textcolor{lightgray}{Real}}\hspace{9.3cm}\textbf{\textcolor{lightgray}{R2D2 Simulated}}}\end{overpic}\vspace{1px}     
        
        
        

        \vspace{4px}
        \caption{\textbf{Visual results:} all images have the real sonar scans on the left, and the R2D2-simulated images on the right. The horizontal pairs of images correspond to the same semantic maps. The miniature image at the top-left corner is the Marine Sonic sonar data, generated with the MC-pix2pix method \cite{MC-pix2pix}. The rest of the images are EdgeTech sonar scans, generated with R2D2-GANs, provided in a relative scale according to the corresponding across track resolutions. The partial overlay in the top-right corner is an example of the semantic map, used by a generator network in order to control the topography of the simulation according to preferences of the user.}
        \label{visuals}
    \vspace{-4px}
    \end{figure*}

The visual examples of the results of the R2D2-GANs are provided in the Figure~\ref{visuals}. Results demonstrated in this work are acquired with the image-to-image translation based architecture \cite{pix2pix2016}, which could be easily altered to accommodate another type of GAN, in order to better facilitate different simulation objectives.

\section{RELATED WORK}
The focus of this paper is the simulation of continuous side-scan sonar imagery of any requested size and resolution, with the user-controlled topography. Because of the visual nature and preferred stochasticity of such simulation, we focus on the corresponding family of the generative models.

Generative Adversarial Networks (GANs) were first introduced in 2014 \cite{vanillaGAN}. Since then, they grew into a highly diverse class of methods and became the most popular way of the realistic image and video generation~\cite{videoGAN}, image completion~\cite{InpaintingGAN}, super-resolution~\cite{SuperGAN}, and style transfer \cite{CycleGAN2017, pix2pix2016, pix2pixHD, SPADE2019}. There has also been a number of alternative applications, such as generation of socially acceptable trajectories \cite{SocialGAN} and control policy generation with GANs~\cite{infoGAIL, GPN}. Nevertheless, the visual data still stays the primary domain of application and development of the GAN models.

Despite that, there is still a relatively limited number of applications of GAN-based methods to the underwater sonar domain. Until recently the only application was the sonar imagery enhancement, mostly for the ATR training purposes - applying CycleGANs-powered style transfer to enhance the synthetic targets for the ATR training sets~\cite{SSPDcycleGAN}, and refining underwater video images~\cite{Fabbri2018ICRA}. However, there is almost no work focusing on the generation of the whole missions worth of the synthetic sonar data with complex terrains.

The first published work to bridge this exception was MC-pix2pix~\cite{MC-pix2pix}. This method produces continuous full-mission-long sonar images for smaller across track resolution sonars (such as Marine Sonic). The results are both indistinguishable from real by human experts and capable of boosting the performance of the ATR systems. MC-pix2pix facilitates realistic conditional generation of the user-specified terrains with a modified pix2pix-style image translation. MC-pix2pix exploits Markov assumption for sequential generation of the image fragments in the along track direction, providing the continuity of the resulting image. An additional advantage of this piece-wise architecture is that it is relatively undemanding about the hardware. It is being supported even for the GPUs with very modest RAM capacities, which is almost never an option for the other higher-resolution GAN types.

Our new method, the R2D2-GAN, retains the general performance level of the MC-pix2pix method and completely surpasses the former on the magnitude of the across track image resolution it is capable of generating.

    \begin{figure}
        \vspace{0.5cm}
        \centering
        \begin{overpic}[scale=0.75]{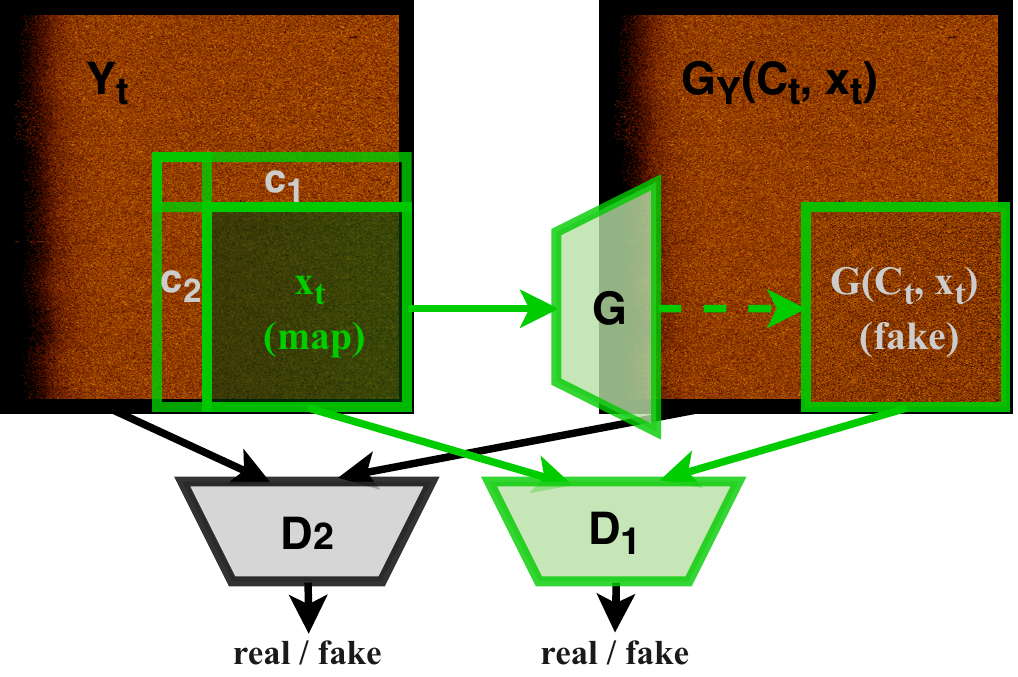}\put(-1.5, 67.5){\textbf{\textcolor{gray}{Training time:}}}\end{overpic}\vspace{4px}

        \caption{\textbf{At training time:} inputs of the generator network include conditions $c_1$ and $c_2$, which are small snippets of adjacent tiles (top and left) of the currently generated tile. The output of the generator $G$ is the suggested synthetic image tile, generated taking into account the input conditions. The output of the generator is then assessed by the first discriminator $D_1$ along with real tiles. The second discriminator $D_2$ assesses the larger real image variants ($2 \times 2$ tiles) - the unchanged and the edited with a generated tile. Both discriminators issue their decisions on whether the image is real or synthetic, and are rewarded based on the correctness of their decisions. Losses of both discriminators are then back-propagated through to the generator and used for the adversarial training.}
        \label{scheme_train}
    \end{figure}

    \begin{figure}
        \centering
        \vspace{0.5cm}
        \begin{overpic}[scale=0.65]{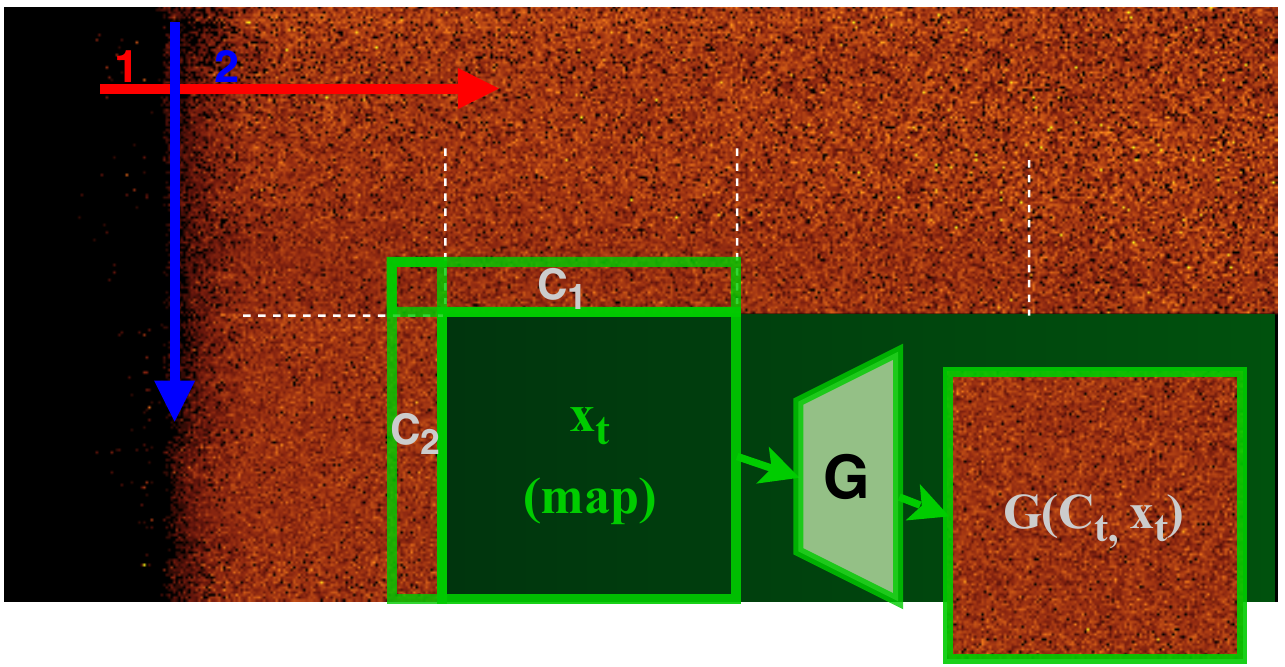}\put(3, 53.5){\textbf{\textcolor{gray}{Test time:}}}\end{overpic}\vspace{4px}

        \caption{\textbf{At test time:} only the generator is used at this stage. It produces image tiles first left-to-right, and then top-to-bottom. Each tile is conditioned on adjacent image snippets of the tile above and to the left of the currently generated tile. These conditions help to maintain the continuity of the larger picture produced at test time.}
        \label{scheme_test}
        \vspace{-3px}
    \end{figure}

\section{METHOD}
The R2D2 belongs to the family of GANs. More specifically the R2D2 is an extension of the Markov-conditional image-to-image translation technique, MC-pix2pix \cite{MC-pix2pix}, re-purposed to enable image generation in larger resolutions.

\keypoint{GANs at a glance:} the default GAN architecture usually consists of two neural networks. The first one, called the discriminator, learns to distinguish the real training images from the synthetic ones, whereas the second one, the generator, trains to create synthetic images that the discriminator cannot distinguish from real ones. Both networks are trained completely from scratch, gradually improving performance in an iterative manner via adversarial training. The final result of the GAN training is usually the trained generator network capable of generating diverse realistic images.

There are multiple conditional flavours of this basic architecture. Henceforth we focus on the paired image-to-image translation techniques, of which the pix2pix~\cite{pix2pix2016, pix2pixHD} is a prime example. This choice is dictated by the requirement of the user-controlled topography.

\keypoint{The main features of the R2D2 technique:} (i) incremental recursive generation (extending the Markov principle from \cite{MC-pix2pix}) applied along two axes (rather than just one, like in \cite{MC-pix2pix}). This allows for handling any across track resolution. (ii) an additional discriminator is introduced for the coherence control of resulting larger scale images. Figures~\ref{scheme_train} and~\ref{scheme_test} provide the schematic illustration of the method.

\keypoint{At training time:} as per scheme in Figure~\ref{scheme_train}, first, the larger training images and their semantic maps are partitioned into multiple tiles. The generator inputs noise and the semantic maps of the current tile - for topography control. It also uses the additional conditions that include the location of a pixel in the across track direction and small snippets taken from the adjacent tiles above and to the left of the current tile. These are to ensure the continuity of the resulting large image. The generator is trained to produce realistic images given the above inputs and conditions.

The discriminator $D_1$ then tries to distinguish the real imagery from the simulated. The discriminator $D_2$ does the same, but processing the larger images ($2 \times 2$ tiles) with the newly generated tiles embedded into them, and tries to distinguish them from the real unedited larger images.

The results provided in this paper are building upon the fully-convolutional pix2pix-style architecture with 9 resnet blocks \cite{pix2pix2016}, extended with additional conditions to support incremental recursive generation and additional ``bigger picture" discriminator $D_2$
 to further encourage the smoothness and continuity of the resulting image.
This model is adversarially trained for 10 epochs with batch-size 3, and 3 gradient updates of discriminator $D_1$
corresponding to each gradient update of the generator. The training loss function can be summarised as follows:

\[\quad\quad\hspace{0.05cm}{G_t}^\star = \arg \min_G \max_D \big\{\mathbb{E}_{\mathbf{C}_t, x_t, y_t, z}[\left\|y_t - G(\mathbf{C}_t, x_t, z)\right\| _1]\]
\[ + \hspace{0.2cm}\frac{1}{2}\Big(\mathbb{E}_{\mathbf{C}_t, x_t, y_t}[log D_1(\mathbf{C}_t, x_t, y_t)] + \mathbb{E}_{X_t, Y_t}[log D_2(X_t, Y_t)] \] 
\[\hspace{0cm} +\hspace{0.25cm} \mathbb{E}_{z, \mathbf{C}_t, x_t}[1 - log D_1(\mathbf{C}_t, x_t, G(\mathbf{C}_t, x_t, z))] \]  
\[\quad\hspace{0.55cm} +\hspace{0.25cm} \mathbb{E}_{\mathbf{C}_t, z, X_t}[1 - log D_2(X_t, G_{Y_t}(\mathbf{C}_t, x_t, z))]\Big)\big\}     \quad(1) \]

where $x_t$ are semantic maps and $y_t$ are real sonar images per tile. $X_t$ and $Y_t$ are larger ($2 \times 2$ tiles) semantic maps and sonar images respectively. $X_t$ and $Y_t$ include tiles $x_t$ and $y_t$ correspondingly. $z$ is a random noise vector, and $\mathbf{C}_t = [c_1, c_2] $ are the condition variables for the generator. $G_{Y_t}(\mathbf{C}_t, x_t, z)$ stands for a larger image $Y_t$ ($2 \times 2$ tiles), where the native tile $y_t$ is replaced by the generated tile $G(\mathbf{C}_t, x_t, z)$. The first line of Equation~(1) represents the L1 loss, a regularization term meant to reduce the blurring in the generator output \cite{pix2pix2016}. The second line stands for the losses of both discriminators classifying the real data, and the last two lines are their losses of classifying the generated data.

\keypoint{At test time:} the trained generator produces the entire image continuously piece by piece, first left-to-right, and then top-to-bottom, in accordance with the requested semantic maps. Refer to the Figure~\ref{scheme_test} for the schematic explanation.

\keypoint{Generalisation:} the underlying technique of the R2D2 can be generalised beyond the specific GAN architecture. Nearly any GAN-based network, depending on the objectives and constraints of a specific task, can in principle be extended for the incremental recursive generation in a manner similar to the R2D2. The results provided in this paper are based on extending pix2pix-style architecture \cite{pix2pix2016} solely for the purpose of providing the user with full control over the topography of the synthesized mission via utilisation of semantic maps.


\section{EXPERIMENTAL SETUP}

\keypoint{Visual quality tests:} a number of assessments were conducted in order to quantify the realism of the obtained imagery. We invited 10 domain experts to evaluate a selection of synthetic images created with the R2D2 and with classic pix2pix (for comparison), along with the real images. Participants inspected these images, labelling them as ``real" or ``synthetic" (``fake"). 

All the image sets of the compared methods were presented in even proportions. Furthermore, all of the image sets (both real and synthetic) correspond to exactly the same set of the semantic maps to ensure the best possible comparability. Images acquired from the different sources were shown sequentially, one at a time, in order to avoid the cognitive bias. For the same reason there was no prior information provided on the proportion of real vs. synthetic images. The only information provided was that the test set contains both real and synthetic images.

Although the time taken to inspect each particular image was recorded and analysed, there was no time constraints imposed on the participants during the test time.

\begin{figure}
    \centering
    
    \begin{overpic}[scale=0.95]{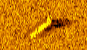}\put(3,40){\textbf{\textcolor{white}{\circled{1}}}}\end{overpic}\hspace{0.06cm}\begin{overpic}[scale=0.685]{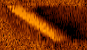}\put(3,40){\textbf{\textcolor{white}{\circled{2}}}}\end{overpic}\hspace{0.06cm}\begin{overpic}[scale=0.685]{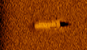}\put(3,40){\textbf{\textcolor{white}{\circled{3}}}}\end{overpic}\hspace{0.06cm}\begin{overpic}[scale=0.685]{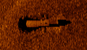}\put(3,40){\textbf{\textcolor{white}{\circled{4}}}}\end{overpic}

    \vspace{0.06cm}
    
    \includegraphics[scale=0.95]{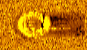}\hspace{0.06cm}\includegraphics[scale=0.685]{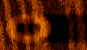}\hspace{0.06cm}\includegraphics[scale=0.685]{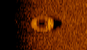}\hspace{0.06cm}\includegraphics[scale=0.685]{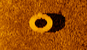}
    \vspace{-5px}
    
    \caption{\textbf{Examples of target objects (tyres and cylinders) and seabed types used for training the ATR:} 1. Uniform random noise background. 2. SonarSim-generated terrains\textsuperscript{1}. 3. R2D2-generated terrains. 4. Real terrains. All of these targets are inserted into the terrains using the Cycle-GAN-based technique \cite{SSPDcycleGAN}, because of the limited availability of the real data with real targets.}
    \label{targets}
    \vspace{-3px}
\end{figure}

\keypoint{Autonomous target recognition (ATR) training:} we argue that our proposed technique does not solely look good to human eye, but also can be of help for the autonomous systems training. For instance, assuming the lack of training data for the ATR, one could boost the training with the R2D2-generated data.
Unfortunately, we do not possess unrestricted real data for sonars in EdgeTech resolution range (4620 or higher across track resolution), that would contain any real objects. Which is why we use the Cycle-GAN-based technique \cite{SSPDcycleGAN} to embed the artificial objects. In our specific case - cylinders and tyres, see Figure~\ref{targets}. In fact, there is a very small amount of unrestricted real seabed scans available, so we have to use what little real data we have for the test set. 

There are a few training sets: (i) uniform random noise background, (ii) SonarSim\footnote{ SonarSim - standard hard-coded and vaguely realistic side-scan simulator as used in \cite{SSPDcycleGAN}, capable of generating various seabed textures with limited user control over the type of generated data, but not the exact topography.} terrains - flat and rippled (respectively easier and harder for ATR to learn to operate on), and (iii) R2D2-generated terrains. Finally, we test the trained ATR performance on the small amount of the real sonar images we have available.

We use a simple ResNet-type architecture for ATR, trained from scratch for this experiment. Both the training and the test sets are rather small, due to the unavailability of the real data. Note that we do not claim the state-of-art level of ATR results here, only the relative benefit of using the R2D2-generated data in the absence of the real training data. 

\begin{table}
\vspace{2px}
\begin{center}
\ra{1.3}
\begin{tabular}{c c c c c }
\hline
Metrics: & fake labelled `real' & accuracy & av. time \\
\hline
\hline
\multicolumn{1}{l}{pix2pix} & 0.14 & 0.88 & 4.79 \\
\multicolumn{1}{l}{R2D2 unnormalized} & 0.26 & 0.82 & 4.80 \\
\multicolumn{1}{l}{R2D2} & \textbf{0.78} & \textbf{0.56} & \textbf{5.23} \\
\hline
\end{tabular}
\end{center}
\vspace{2px}
\caption{\textbf{Visual test results:} the experiment was conducted with participation of 10 human experts possessing the daily experience of dealing with the sonar imagery. They were shown an equal number of images generated by different sources (both simulated and real) and asked to label them as ``real" or ``fake". Images coming from different sources were shown one after the other in a random order to mitigate the cognitive bias. \newline
R2D2 images were labelled ``real" in 78\% cases, which compares well with the benchmarks, as well as with real images labelled ``real" (90\%). Humans also were able to distinguish it from real with accuracy of 56 \%, which is close to random chance in a two-class problem (``real" / ``fake"). The last column of the results shows how long (in seconds) on average it took to classify an image. R2D2-produced images took significantly longer to process, compared to the other methods.}
\label{results}
\end{table}



\section{RESULTS}
\keypoint{The visual results of the R2D2-GANs} are shown in Figure~\ref{visuals}, there is very little to no difference between the real (left) and synthetic (right) images. Also, please note the relative difference in scales between the typical data generated with MC-pix2pix (top-left corner) and R2D2-GANs (right). Because of the iterative recursive generation along both axes, R2D2 is practically unlimited in the data resolution it is able to generate. Naturally, there is always a trade-off between the magnitude of the generated image resolution and the generation speed.

\keypoint{The image assessment scores by human experts,} presented in the Table~\ref{results}, are based on the results collected from 10 human experts with various level of expertise, but dealing with sonar imagery on a daily basis. The individual assessment metrics are as follows:

(i) R2D2 synthetic imagery is labelled ``real" by humans in 78$\%$ of the cases. This is the highest score across the competing methods, which also compares reasonably well to 90$\%$ score for the real images classified as real.

(ii) Human classification accuracy being close to 50$\%$, i.e. near random chance for two-class problem (``real'' / ``fake''), indicated inability of humans to tell apart the real and synthetic images. The R2D2 shows the lowest human classification accuracy score of 56$\%$. This is comparable with the 52$\%$ score obtained by the MC-pix2pix in an identical experiment \cite{MC-pix2pix}, which is a remarkable result considering the significantly higher complexity of the current task of the higher-resolution generation.

(iii) We do not attribute any definite meaning to the average time spent on inspection of each separate image. Nevertheless, images produced by R2D2-GANs take the longest to classify. We suggest to interpret this as the R2D2-generated synthetic imagery posing the higher challenge for distinguishing it from real.

Our method outperforms the original pix2pix according to all the metrics in this assessment. It is also comparable with the current state of art - MC-pix2pix~\cite{MC-pix2pix}, which achieves the human labelling accuracy of 52$\%$ for images of much smaller resolutions. The R2D2, however, surpasses this competitor by the resulting resolution of the complete images it is capable of generating, while maintaining the comparable quality of the generated results.

\setlength{\tabcolsep}{0.42em}
\begin{table}
\vspace{2px}
\begin{center}
\ra{1.3}
\begin{tabular}{c c c c c }
\hline
Train set: & Noise & SonarSim (flat)& SonarSim (rippled)& R2D2-GAN \\
\hline
\hline
\multicolumn{1}{l}{Recall} & 0.00 & 0.3314 & 0.2255 & \textbf{0.4843} \\
\multicolumn{1}{l}{F1} & 0.00 & 0.4895 & 0.3653 & \textbf{0.6073} \\
\hline
\end{tabular}
\end{center}
\vspace{2px}
\caption{\textbf{Autonomous Target Recognition experimental results:} unfortunately, we have no access to unrestricted data with real targets at our disposal. However, we demonstrate, that expanding the ATR training datasets with R2D2-generated data may help the learning process. Potentially, a higher variety of the terrains available at training time should help the ATR system to generalise better. Fortunately, there is a Cycle-GAN-based method \cite{SSPDcycleGAN} to insert some artificial objects into the terrains, that is useful in this case. The training is conducted on 4 different types of terrains: uniform random noise, SonarSim\textsuperscript{1} flat and rippled terrains (respectively less and more challenging for the ATR), and the R2D2-generated terrains. We train a simple ResNet-type network over these, and test on the real data with artificial targets embedded. \newline
The results suggest that random noise in this case is completely useless, whereas R2D2-GAN on the contrary performs better than the competitors.}
\label{atr}
\end{table}

\keypoint{ATR performance results} are available in Table~\ref{atr}. Both recall and F1-score\footnote{ F1-score -  harmonic mean between the precision and the recall of the ATR system. Higher values correspond to the better performance.} suggest that the R2D2-generated terrains provide significantly better training material compared to the random noise and SonarSim simulator.

\section{CONCLUSIONS}
This paper presents the R2D2-GANs - a novel technique for generating the realistic synthetic imagery of any specified resolution and topography. This work provides both the quantitative and qualitative evidence confirming the realism of the images produced with our method. The empirical assessment also suggests significant advantages for the ATR systems trained with R2D2-generated data.

The presented technique is in principle compatible with nearly any type of GANs, which might be of benefit for alternative objectives than those explored in this work. Thus providing the user with the ultimate control over the exact nature of the preferred data generation process. 
The R2D2-GANs are practically unlimited in the image resolution they can generate (at expense of the generation speed), which makes them immediately applicable to even higher resolution sonars, such as the Synthetic Aperture Sonars. Nonetheless, in the future work we intend to optimise this method further to make sure the fastest possible speed of the image generation for even higher resolutions.

\section{ACKNOWLEDGEMENTS}
We would like to express our gratitude to Stephanos Loizou for his help with the ATR experiments, as well as the employees of the Seebyte Ltd who were of great help in gathering the data for the Table~\ref{results} by participating in the image assessment tests. 





\bibliographystyle{IEEEtran}
\bibliography{IEEEabrv,IEEEexample}

\end{document}